%% file: 00_starthere.tex
\documentclass[11pt]{article} 

\usepackage{bm} 
\usepackage{color} 
\usepackage{hyperref} 
\usepackage{graphicx} 
\usepackage{vmargin} 
\usepackage{afterpage}

\usepackage{pdfpages}

\definecolor{green}{RGB}{11,155,13}

\hypersetup{colorlinks=true,linkcolor=blue,urlcolor=blue}
\setpapersize{USletter}
\setmarginsrb{1in}{1in}{1in}{1in}{0pt}{0mm}{0pt}{0mm}

\newcommand{\pseudodot}{{\lower 2.4pt\hbox{$\cdot$}}}
\usepackage[utf8]{inputenc}
 
\begin{document}

\setlength{\baselineskip}{14pt} 
\setlength{\normalbaselineskip}{12pt} 

\section*{Collaborative Design for Job-Seekers with Autism:\\A Conceptual Framework for Future Research}

Sungsoo Ray Hong\textsuperscript{\textdagger},
Marcos Zampieri\textsuperscript{\textdagger},
Brittany N. Hand\textsuperscript{\textdaggerdbl},
Vivian Motti\textsuperscript{\textdagger},
Dongjun Chung\textsuperscript{\textdaggerdbl},
and
Ozlem Uzuner\textsuperscript{\textdagger}
\footnote{
Dr. Sungsoo Ray Hong is an Assistant Professor in Information Sciences \& Technology at George Mason University. He specializes in Human-Computer Interaction (HCI) and Computer-Supported Cooperative Work (CSCW) (\href{mailto:shong31@gmu.edu}{shong31@gmu.edu} \url{http://www.rayhong.net/}). 
Dr. Marcos Zampieri is an Assistant Professor in Information Sciences \& Technology at George Mason University. His domain of research is computational linguistics and Natural Language Processing (\href{mailto:mzampier@gmu.edu}{mzampier@gmu.edu} \url{https://www.gmu.edu/profiles/mzampier}).
Dr. Brittany N. Hand is an Associate Professor in Health Outcomes and Injury Prevention Sciences at the Ohio State University.  Her expertise includes health services and primary care for autistic adults 
(\href{mailto:brittany.hand@osumc.edu}{brittany.hand@osumc.edu} \url{https://hrs.osu.edu/faculty-and-staff/faculty-directory/hand-brittany}).
Dr. Vivian Motti is an Associate Professor in Information Sciences \& Technology at George Mason University. Her research domain includes HCI and accessibility computing (\href{mailto:vmotti@gmu.edu}{vmotti@gmu.edu} \url{https://volgenau.gmu.edu/profiles/vmotti}). 
Dr. Dongjun Chung is an Associate Professor in the Department of Biomedical Informatics at the Ohio State University. His research focuses on developing statistical and computational models for the analysis of biomedical big data (\href{chung.911@osu.edu}{chung.911@osu.edu} \url{https://sites.google.com/site/statdchung/}). 
Dr. Ozlem Uzuner is a Professor of Information Sciences \& Technology at George Mason University. She specializes in Natural Language Processing, Artificial Intelligence, and Machine Learning (\href{mailto:ouzuner@gmu.edu}{ouzuner@gmu.edu} \url{https://volgenau.gmu.edu/profiles/ouzuner}).
}

\vspace{2mm}
\noindent\textsuperscript{\textdagger}George Mason University, Fairfax, VA, USA\\
\noindent\textsuperscript{\textdaggerdbl}Ohio State University, Columbus, OH, USA

\paragraph{Abstract}
The success of employment is highly related to a job seeker's capability of communicating and collaborating with others.
While leveraging one's network during the job-seeking process is intuitive to the neurotypical, this can be challenging for people with autism.
Recent empirical findings have started to show how facilitating collaboration between people with autism and their social surroundings through new design can improve their chances of employment. This work aims to provide actionable guidelines and conceptual frameworks that future researchers and practitioners can apply to improve collaborative design for job-seekers with autism.
Built upon the literature on past technological interventions built for supporting job-seekers with autism, we define three major research challenges of (1) communication support, (2) employment stage-wise support, and (3) group work support.
For each challenge, we review the current state-of-the-art practices and possible future solutions. We then suggest future designs that can provide breakthroughs from the interdisciplinary lens of human-AI collaboration, health services, group work, accessibility computing, and natural language processing.

\input{sections/01_introduction}
\input{sections/02_relatedwork}
\input{sections/03_problemspace}
\input{sections/04_conclusion}
\input{sections/05_acknowledgement}

\bibliographystyle{IEEEtran}
\bibliography{sections/99_ref}

\end{document}

%% file: sections/01_introduction.tex
\section{Introduction}

In April 2023, the employment rate in the US was historically low at 3.4\%.
At the same time, more than 85\% of people with autism were unemployed~\cite{mydisab2024}.
Past studies have largely focused on understanding how employers can implement more neurodiversity-inclusive hiring environments and open workplace culture~\cite{neuro1, emerg, inproceedingsweli}. 
However, in 2024, it is estimated that over 5 million people with autism will be unemployed in the US~\cite{CDC2024}.
Unemployment can reduce the quality of life among people with autism in many ways, for example, by preventing independent living, isolating them from peer groups and communities, or contributing to poor mental health~\cite{taylor2022job, Hedley_Cai}.

Collaborative design is an emerging direction of research that aims to lead to better outcomes for job seekers with autism by facilitating collaboration between them and their social surroundings~\cite{ara2024collaborative}.
When people seek a job, they naturally collaborate with their social surroundings---colleagues, family, friends, career advisors, employers, and other relevant parties~\cite{ZIKIC2009117,  MARTIN2021101741, Westbrook}.
Well-shaped communication and organized collaboration with their social surroundings are crucial for success in job seeking~\cite{Liu2014-xs, schwarzer2007functional}.
However, job-seekers with autism may have disadvantages in collaborating with others compared to the neuromajority~\cite{dillahunt2021implications}.
One possible reason can be a different communication style; people with autism are ``literal thinkers'' who minimize using contextual language~\cite{ara2024collaborative}.
For example, someone with autism may have difficulty answering a job interview question of ``Where do you imagine yourself in 3 years?'' as they may not know whether they will get the job, and if so, whether they will still work there in 3 years~\cite{ara2024collaborative}.
Other reasons may include executive dysfunction~\cite{hill2004executive} and emotional dysregulation~\cite{kalantari2021emotion}.

Literature on collaborative job-seeking for people with autism has predominantly focused on gaining an empirical understanding of their current practice and challenges while having relatively less emphasis on devising new interventions to provide better strategies. 
To provide actionable guidelines and future research directions in the problem space of collaborative design with job seekers with autism, we synthesize the existing findings in the literature and define major research challenges using interdisciplinary viewpoints that combine human-AI collaboration, group work, natural language processing, clinical health services, and beyond. 
Consequently, we define challenges in the conceptual framework of Collaborative Design for Job Seekers with Autism as follows:
\begin{itemize}
    \item \textbf{Communication Support Challenge}: The language of people with autism is more likely to be literal compared to neurotypical peers. How can we leverage technology to help them decode various linguistic devices hidden in neurotypical language, such as euphemism, sarcasm, simile, structure mapping, metaphor, and context, when reading and listening?
    \item \textbf{Employment Stage-wise Support Challenge}: Arguably, the current job-seeking support tools have been built from the perspectives of neurotypicals. Such design cannot fully accommodate the needs of people with autism in specific stages of the hiring process, such as reading job calls, resume preparation, or interviews. How can we redesign technology to provide autism-congruent solutions that reduce cognitive burdens and facilitate job-seeking in collaboration with their social surroundings efficiently and effectively?
    \item \textbf{Group Support Challenge}: When working in groups, handling the communication load, synchronizing situational awareness and knowledge among members, and making consensus-driven decisions can be hard for people with autism. Such difficulties can get harder in multi-tasking scenarios such as job-seeking. In supporting the collaborative effort between people with autism and their social surroundings, our framework's last challenge is understanding how to facilitate group collaboration through intelligent design.
\end{itemize}

In elaborating on each challenge space, we start by synthesizing the existing work to define a set of possible directions of research that correspond to an application-driven challenge.
Based on our multidisciplinary expertise, we provide our viewpoint on how different domains of researchers can collaborate to make progress.
This work provides opinion contributions across the fields of Human-Compouter Interaction (HCI), Computer-Supported Cooperative Work (CSCW), Natural Language Processing (NLP), Autism research, and Health Service explaining how they can practically support job-seekers with autism by providing novel future solutions.


%% file: sections/02_relatedwork.tex
\section{Related Work}
\label{sec:RW}

People with autism can meaningfully contribute to their workplace with several skills, such as hyperfocus and consistency~\cite{nicholls2023autistic}.
They can pay close attention to critical details others often miss, think outside the box, and are more likely to report and address problems that could otherwise be unaddressed~\cite{cope2021strengths}.
Despite such potential, the census shows that the unemployment rate for adults with autism was over 80\% for a decade~\cite{beth2014unemployment, mydisab2024}.
Since employment is essential for maximizing the quality of life and independence for people with autism, research has focused on building autism awareness and acceptance in the workplace or providing accommodations to lower employment barriers~\cite{10.1371/journal.pone.0147040}.
For instance, Lorenz et al. found the lack of sensitivity and awareness about workers on the autism spectrum can introduce a ``formality problem'' at an organizational level and impede workers with autism from being productive~\cite{lorenz2016autism}.
Annabi and Locke reviewed the literature on the employment of people with autism and developed a theoretical framework for assessing IT workplace  readiness~\cite{annabi_locke_2019}.

While there is a ``macroscopic'' approach to reducing unemployment among adults with autism by increasing awareness and neurodiverse-friendly workplace culture~\cite{ara2024collaborative}, other studies attempted to build new technological interventions---though we have ``limited volume'' of research and unestablished ``quality of methodology''~\cite{nicholas2015vocational}. 
For example, one review study conducted in 2014 found only 10 articles focused on employment-related intervention for adults with autism over 18 years old among 3,874 autism intervention articles~\cite{nicholas2015vocational}.
While there is relatively less volume of studies for adults with autism, empirical studies found people with autism transitioning from secondary education to the workplace are in a precarious position due to a dramatic reduction of formal supports and services~\cite{ezerins2023autism}. Motivated by such findings, more researchers committed efforts to developing better technological interventions for reducing barriers for job seekers with autism.  For example, a survey conducted in 2022 found 48 studies that propose employment-based interventions~\cite{Kim_Crowley_Lee_2022}.

Collaborative design is an emerging direction of intervention for facilitating autistic employment through collaboration with their social surroundings~\cite{ara2024collaborative}.
Existing employment interventions for adults with autism have leveraged social support, such as by job seekers eliciting feedback, guidance, or obtaining job-related video tips from volunteers~\cite{Kim_Crowley_Lee_2022, Interview4}.
However, we have identified the three gaps we aim to address in our framework. 
First, social collaboration has a variety of formations---synchronous or asynchronous, and distributed or co-located. Additionally,  activities entailed in job-seeking can be highly diverse, including writing e-mails, deciding where to apply, preparing for interviews, or emotionally recovering from the arduous process of job-seeking. 
While approaches exist that apply social support in employment intervention, collaborative design has not been systemically and comprehensively analyzed in terms of how it can benefit people with autism in varying job-seeking situations. 
In developing the scope of collaborative design, we synthesize the literature and define the major directions that can create synergy between people with autism and their social surroundings. 
Second, job-seeking is a longitudinal effort that requires consistent and recurring support. 
Finding volunteers who can commit such consistent help can be challenging. 
To make future endeavors more practical, our framework confines the social surroundings as two types: (1) family or close colleagues of the job-seeker with autism, or (2) professionals working with job-seekers with autism, such as occupational therapists or career advisors. 
Third, we are witnessing the burgeoning growth of interest in applying AI to various tasks. 
While extensive conversation has started to reveal how to benefit various user groups using AI, we lack investigation regarding how AI, such as Large Language Models (LLMs), can benefit people with autism and their social surroundings when conducting job-seeking-related activities.




%% file: sections/03_problemspace.tex
\section{Research in Collaborative Job-seeking for People with Autism}

In laying out the main directions in our framework, we collected noteworthy findings from three areas of literature: (1)  reviews of interventions for job-seekers with autism (e.g., ~\cite{ezerins2023autism, khalifa2020workplace, Kim_Crowley_Lee_2022, nicholas2015vocational}), (2) empirical studies that provide insights into how people with autism leverage technology in seeking jobs (e.g., ~\cite{dillahunt2021implications, dillahunt2016designing, ara2024collaborative}), and (3) technical papers that propose interventions to improve job-seeking experiences for people with autism, with or without social supports(e.g.,~\cite{adiani2022career, pouliot2017tool} ).
In our analysis, the first author coded the papers regarding any noteworthy empirical insights and opinions. 
Using these codes as an initial resource, every author participated in affinity diagramming and theme-generation process following qualitative research~\cite{saldana2015coding, layder1998sociological}. 
In shaping our final themes, each of the authors built their structure based on their expertise and compared the structures to identify commonalities and discrepancies
The discussion continued multiple rounds until every author reached a consensus. 
Consequently, we identified the three directions---Communication Support, Employment Stage-wise Support, and Group work support---as imminent and important problems that can yield practical impact on the collaborative design for job-seekers with autism. 

\subsection{Communication Support} 

\paragraph{Background and State-of-the-Art}

One of the most common challenges discussed in the literature is the different communication styles between people with autism and neurotypicals. 
In particular, people with autism tend to be literal thinkers whereas neurotypical language uses various implicit language devices, such as abstraction, metaphor, mapping, sarcasm, or euphemisms.
Past research has identified such bidirectional communication breakdown as a core reason for people with autism to be not successful in the employment market~\cite{lorenz2016autism, ara2024collaborative}
Some communication challenge examples featured in the previous literature include: interpreting implicit language such as metaphors, analogies, and idioms in the intended way; handling casual topics (i.e. small talk unrelated to the job search); adjusting language and level of formality depending on the conversational context; and taking turns (i.e., balancing conversational flow between two or more people).
A recent study found how different styles of communication can cause unsuccessful outcomes in every stage of job-seeking; when reading job calls, asking for details about jobs to hiring managers, conducting interviews, and negotiating the offer~\cite{ara2024collaborative}.

Mitigating the bidirectional communication gap is the core to building up social skills related to their career~\cite{ezerins2023autism}.
Therefore, it is imperative to devise technological intervention to mitigate the gap between literal language and contextual language ~\cite{nicholas2015vocational}. 
Such language translation is relatively scarce in literature but is emerging.
For instance, Pouliot et al. devised a communication supporting method~\cite{pouliot2017tool}. 

\paragraph{Research Directions}

In developing a new intervention for conversational support, the requirements will be different between the following two cases of \textit{conversational decoding}---people with autism understanding neurotypical language---and \textit{encoding}---people with autism using neurotypical language. In decoding, it is crucial to develop ways to detect the implicit use of language and translate it to the explicit language. Recent advances in generative AI, more specifically LLMs combined with prompt engineering, can be applied in developing encoding scenarios. In encoding, language style transfer and tone changing can be useful. One aspect of communication that is difficult for people with autism is their anxiety about being perceived as rude or being misunderstood~\cite{ara2024collaborative}.
In many cases, people with autism get help from their social surroundings to check the ``tone and manner'' of their writing style~\cite{ara2024collaborative}. 
While social surroundings can proofread writing and guide people with autism to formulate answers to expected interview questions, providing such feedback across the whole period of the job-seeking process can be time and resource intensive. 
Therefore, designing intelligent interventions that can support social surroundings will be perceived as useful. Finally, we note that such decoding and encoding support will be equally important and useful for neurotypicals in their collaboration.

In developing such intelligent group-AI collaboration interfaces in both directions of conversation, we list a few more related challenges. To build AI models that can be specifically accurate for conversational encoding and decoding, characterizing the common cases of misalignment using numerous cases of conversation made in employment contexts is essential. Such case collection and classification are under the domain of data annotation. There has been extensive research in interactive data annotation~\cite{choi2019aila}. However, few approaches have been proposed for making the annotation efficient and effective in job-seeking scenarios for people with autism.
In addition, in developing interactive user interfaces used for conversational encoding and decoding, strictly applying a human-centered and/or participatory methodology by closely communicating with multi-stakeholders involved (i.e., people with autism and various social surrounding groups) can increase the chance of building the intervention that is perceived as practical, useful, and be adopted in situ to create impact. 

\subsection{Employment Stage-wise Support}

\paragraph{Background and State-of-the-Art}
Employment is a process with multiple stages.
While such a process can be challenging for everyone, numerous articles explain why and how each stage has greater barriers for people with autism than their neurotypical peers~\cite{ara2024collaborative}, from reading a job advertisement to negotiating an offer.
A few empirical studies identified the importance of step-by-step, end-to-end support such as writing cover letters, resumes, and intelligent tutoring~\cite{dillahunt2021implications}.

In terms of state-of-the-art supports and interventions, \textit{DreamGig} helps underserved job-seekers find job openings that align with their career goal and supports them in identifying skills they need to achieve to reach their dream jobs~\cite{dreamgig}.
there have been another line of interactive systems built for supporting employment stages, such as \textit{Review-Me} that supports resume editing and \textit{Interview4} that elicit feedback on interviews from volunteers~\cite{dillahunt2021implications}.
While there are not many technological interventions specifically focusing on job-seekers with autism as a user group, past studies' results imply developing a dedicated feature specifically focusing on stage-wise support may improve the performance of job-seekers with autism.

\paragraph{Research Directions}

Literature shows how people with autism can have difficulties in regulating their attention and prioritizing employment-related tasks~\cite{ara2024collaborative}. 
Several challenges occur during the employment process, those are related to the way in which the job is presented, regarding the description of the position and the required qualifications of the candidate. Other issues emerge due to the numerous steps required to prepare the materials, manage the scheduling, and go through a selection process as detailed below.

\textbf{Searching job ads}:
First of all, it is crucial to help people with autism to easily narrow down job calls based on what they are looking for.
There exists a misalignment between what people with autism commonly look for and what current ``go-to'' search tools are providing.
To people with autism, environmental factors, such as noise level, team-related activity factors, such as the frequency and intensity of communication for finishing up their main responsibility, and the organization's culture about Diversity, Equality, and Inclusion (DEI)---either they put the DEI statement in their job ads as ``boilerplate'' or the statement truly represent their culture and a company can work flexibly with their employees to accommodate their needs---can be crucial factors~\cite{khalifa2020workplace}. 
Additionally, some positions require drastic life changes, such as re-location. However, a change in routine is particularly challenging for many people with autism.
While many search tools---or the majority of job advertisements---don't provide these factors, one aspect that future design can consider is to identify how to efficiently categorize numerous job calls based on such factors.
The factors that job-seekers with autism are looking for can be highly individualized and may evolve over time.
Interactive searching environments that flexibly accommodate their ideal job criteria can significantly reduce their effort. 

\textbf{Reading job ads}:
Developing a reading environment that minimizes the chance of long reading of text, which may be challenging for some people with autism, can provide aid. 
Job ads often describe roles and responsibilities purposefully vague to invite a large number of candidates.
They also provide a long text to cover every aspect of the new job opportunity.
That, summed with people with autism tending to be ``literal thinkers'', may result in hesitancy to apply if they feel like they do not sufficiently meet all job requirements.
In developing an autism-friendly environment, recent advances in large AI models can provide several solutions that can meaningfully improve job seekers with autism's job search-related productivity. 
For instance, modern summarization and simplification models can convert job ads to less lengthy forms written in language that will be easier for people with autism to understand.
Text-based Question Answering techniques can provide aid by helping people with autism to comprehend job ads step-by-step through making questions subsequently. comprehend job ads step-by-step by making subsequent questions using their language.

\textbf{Writing cover letter, resume, and preparing materials}:
Finetuning cover letters and resumes can be also challenging for job-seekers with autism especially because they may likely apply for multiple jobs simultaneously. 
An interactive co-editing design between people with autism, an AI agent, and their social surroundings that can easily convert their findings from job ads reading into cover letters and resume editing can become useful in such a scenario.
In addition, the preparation of the job materials can be overwhelming for job seekers with autism.
Unfortunately, in many cases, the required materials for application are not presented in a structured way with templates.
Additionally, not describing the expected contents and specific steps entailed in the hiring process incurs anxiety during the application process.
Not only do they face challenges preparing and submitting the documents required, but also contact the HR personnel to seek help and interview through calls or in-person.
Designing, developing, and evaluating intelligent agents who can support people with autism and their social surroundings to navigate these steps can help job seekers with autism to be better prepared and more successful in their efforts.

\textbf{Conducting interviews}:
For job seekers with autism, participating in the interview is one of the most challenging steps in the hiring process.
Some candidates may also face challenges due to required improvisation when the interview questions are not known upfront.
For them, it can be hard to project themselves into situations that are not yet experienced (imagined scenarios).
On top of the unanticipated questions, making conversation with an unknown group of people adds to the challenge.
In the case where a script with interview questions is not available, developing the mock interview or shadow interview environment that enables (1) people with autism to be exposed to a variety of questions related to the job while (2) social surroundings can easily review and provide their comments can help them.
Also,  people with autism may benefit from practicing conversational turn-taking in the interview context, with emphasis on estimating the right duration and depth of answers to fit within the interview timeframe. 
Moreover, for the candidate to feel comfortable, they may benefit from prompting to take regular breaks.
In helping support the aspects mentioned above, current research in using wearable devices that can measure anxiety levels in real-time or applying devices that provide immersive environments can provide practical future interventions. Gamification design blended with multimodal guidance of providing unique visual and audio feedback depending on how people with autism answer the question or providing cues for time-related information can make a difference.

\textbf{Networking and communication with employers:}
Finally, devising a future networking design that people with autism can leverage in reaching out to volunteers can be helpful. 
However, volunteering in a job-seeking situation expects longitudinal and heavy commitment.
The networking feature should carefully consider why both parties can readily agree to work together. 
In doing so, recent research shows that people with autism have a strong preference for seeking people who have commonalities in the neurodiverse spectrum and cultural fit~\cite{ara2024collaborative}.


\subsection{Groupwork Support}

\paragraph{Background and State-of-the-Art}

Groupwork is the subdomain of CSCW where the aim is to understand how a group of people work together on a certain task using technology in different modes of situations; working synchronously or asynchronously, and at the same space or in a distributed manner~\cite{cook2005illuminating}. 
As one of the most social beings on Earth~\cite{kraut2002aristotle}, humans can collaborate on almost any task under the shared vision. 
While one could imagine that groupware can aid groups easily, Grudin's early work in CSCW has identified that boosting a group's productivity using socialware is extremely challenging~\cite{grudin1988cscw, grudin1994groupware}.
Despite the challenges entailed in designing successful groupware, decades of research have found how groupwork support can create a tremendous impact on several domains that would have been not possible otherwise~\cite{rama2006survey}, enabling advancing democracy, motivating DEI, regulating disinformation, realizing international scientific collaboration, and more~\cite{thaw2008cope, hong2018collaborative}.

As we briefly mentioned in Section 2, we found few studies that systemically leverage groupwork as an intervention to facilitate the job-seeking process for people with autism. 
Meanwhile, recent studies started to show evidence of how groupwork can be a promising direction as a new intervention. 
For instance, using better technology, one study found how underresourced job-seekers can make a difference in their outcomes with career advisors~\cite{dillahunt2021implications}.
With a more productivity-centered groupware design, career specialists can support more people who require help~\cite{lorenz2016autism}. 
This may contribute to making professionals more available to a broader population of autism who could receive benefits~\cite{dillahunt2021implications}.
Another study found varying contexts that tell us how people with autism collaborate with their social surroundings, describing various technological barriers that block them from collaboration~\cite{ara2024collaborative}.

\paragraph{Research Directions}

Supporting group collaboration using technology is challenging because any matters that happen between humans are context-dependent and change every moment.
Since technology is especially effective for handling similar and repetitive activities, Ackerman coined his notion of ``social-technical gap'' which explains why we may never be able to resolve problems that require social collaboration using technology~\cite{ackerman2000intellectual}.
In recent years, advances in large language models and intelligent agents demonstrated meaningful breakthroughs.
Therefore, this section will explain the directions of AI-driven collaboration support that could lead to a successful collaborative design for job seekers with autism.

\textbf{AI agents that help executive planning and multi-tasking}:
One of the most challenging aspects of job-seeking for people with autism is to cope with handling multiple job opportunities that have strict deadlines. 
Once they focus on one task, they may easily fall into time-blindness and be late on the interview or miss deadlines~\cite{ara2024collaborative}.
Handling the deadlines, and helping the group to understand what will come and what needs to be done is a complex task that intelligent agents can support human groups. 

\textbf{AI agents that can soften between-member dynamics and relationships}:
Conflict resolution is one of the most challenging aspects in groupwork~\cite{miranda1993impact}.
A variety of conflicts can arise while working on a high-stakes task with impact, such as job-seeking. 
There are several reasons that conflict can arise; caring for other members can be seen as meddling, and asking for information on each member's activity can be perceived as violating privacy. 
Managing disagreement on the decision can be also an important topic~\cite{harris2018small}. 

\textbf{AI agents that manages the volume of communication and coordinates dependencies for reduced work}:
Finally, the core competency of groupware is to enable the group to achieve their goal in a high quality using reduced effort. 
One downside of groupwork is the necessity of communicating with others to synchronize their situational and group awareness~\cite{hong2018collaborative}. 
In many cases, they must coordinate their group work to manage dependencies between activities~\cite{malone1994interdisciplinary}.
Understanding how to effectively resolve such activities using the AI-empowered agent can create useful solutions.
One of the most challenging aspects of job-seeking for people with autism is to cope with handling multiple job opportunities that have strict deadlines. 
Once they focus on one task, they may easily fall into time-blindness and be late on the interview or miss deadlines~\cite{ara2024collaborative}.
Handling the deadlines, and helping the group to understand what will come and what needs to be done is a complex task that intelligent agents can support human groups.

%% file: sections/04_conclusion.tex
\section{Conclusions}

In this work, we proposed a conceptual framework of Collaborative Design for Job-seekers with Autism that future researchers and practitioners can leverage in supporting people with autism in the job market.
We discussed how recent advances in HCI, CSCW, NLP, Autism Research, Health Service Research, and neighboring areas can be applied to support individuals in three major challenges: (1) communication support, (2) employment stage-wise support, and (3) group support. We hope the challenges and opportunities analyzed here can trigger meaningful technological intervention that can improve the independent living and quality of life of people with autism.

%% file: sections/05_acknowledgement.tex
\section{Acknowledgements}

The authors express gratitude for generous support from the National Science Foundation, Future of Work Grant No.2026513.